**Title**: Photonic topological valley-locked waveguides

*Qiaolu Chen, Li Zhang, Qinghui Yan, Rui Xi, Hongsheng Chen\*, and Yihao Yang\**

Q. Chen, L. Zhang, Q. Yan, Dr. R. Xi, Prof. H. Chen, Dr. Y. Yang
Interdisciplinary Center for Quantum Information, State Key Laboratory of Modern Optical Instrumentation, College of Information Science and Electronic Engineering, Zhejiang University, Hangzhou 310027, China.
(\*hansomchen@zju.edu.cn (Hongsheng Chen); yangyihaooo@zju.edu.cn (Yihao Yang))

Q. Chen, L. Zhang, Q. Yan, Dr. R. Xi, Prof. H. Chen, Dr. Y. Yang
ZJU-Hangzhou Global Science and Technology Innovation Center, Key Lab. of Advanced Micro/Nano Electronic Devices & Smart Systems of Zhejiang, Zhejiang University, Hangzhou 310027, China.

Q. Chen, L. Zhang, Q. Yan, Dr. R. Xi, Prof. H. Chen, Dr. Y. Yang
International Joint Innovation Center, ZJU-UIUC Institute, Zhejiang University, Haining 314400, China

Dr. Y. Yang
Division of Physics and Applied Physics, School of Physical and Mathematical Sciences, Nanyang Technological University, 21 Nanyang Link, Singapore 637371, Singapore.

Dr. Y. Yang
Centre for Disruptive Photonic Technologies, The Photonics Institute, Nanyang Technological University, 50 Nanyang Avenue, Singapore 639798, Singapore.

**Abstract:**

Topological valley kink states have become a significant research frontier with considerable intriguing applications such as robust on-chip communications and topological lasers. Unlike guided modes with adjustable widths in most conventional waveguides, the valley kink states are usually highly confined around the domain walls and thus lack the mode width degree of freedom (DOF), posing a serious limitation to potential device applications. Here, by adding a photonic crystal (PhC) featuring a Dirac point between two valley PhCs with opposite valley-Chern numbers, we design and experimentally demonstrate topological valley-locked waveguides (TVLWs) with tunable mode widths. The photoinc TVLWs could find unique applications, such as high-energy-capacity topological channel intersections, valley-locked energy concentrators, and topological cavities with designable confinement, as verified numerically and experimentally. The TVLWs with width DOF could be beneficial to interface with the exsisting photonic waveguides and devices, and serve as a novel platform for practical use of topological lasing, field enhancement, on-chip communicaitons, and high-capacity energy transport.

# 1. Introduction:

Inspired by the valleytronics,[1-4] researchers have recently proposed a novel class of photonic crystals (PhCs) featuring non-zero Berry curvatures at two opposite valleys in their band structures, also known as valley PhCs.[5-15] At domain walls between two valley PhCs with opposite valley-Chern numbers, valley-locked kink states exist, which are topologically protected against random disorders or defects, due to the strongly suppressed inter-valley scattering. Such robust valley kink states have shown excellent potentials in many modern photonic devices, such as topological channel intersections,[7,14,16] topological lasers,[17] and robust on-chip communications.[11-12,15]

The widths of conventional waveguides, such as silicon rib waveguides and rectangular waveguides, are usually adjustable, providing a new degree of freedom (DOF) for manipulating guided modes and designing photonic devices.[18-21] More remarkably, the waveguides with the width DOF are more flexible to interface with other photonic waveguides or devices than those without the width DOF. However, as the valley kink states are usually highly confined around the domain walls, the previous domian wall structures lack the width DOF, which possess an obvious limitation to potential device applications.

To overcome the above challenges, we design and experimentally demonstrate a topological valley-locked waveguide (TVLW), where a PhC featuring a Dirac point is sandwiched by two valley PhCs with opposite valley-Chern numbers. The topological guided modes in TVLWs are characterized with gapless dispersions, valley-momentum locking, robustness against defects, and, more importantly, a tunable mode width. Such

TVLWs with the width DOF could be very promising in applications. For example, by abruptly adjusting the waveguide width, valley-locked reflection-free energy squeezing effect (similar to the squeezing effect in $\varepsilon$-near-zero narrow waveguide channels)[22-23] could be achieved, which may lead to potential applications in light harvesting[24], field enhancement,[24] and boosting nonlinear effects.[25] Additionally, topological cavities based on TVLWs have tunable mode confinement, which could enable novel topological lasing functions.[17,26-28] Via direct near-field measurements, we experimentally demonstrate a high-energy-capacity topological channel intersection, a valley-locked topological energy concentrator, and a topological cavity with tailorable mode confinement.

## 2. Results and Discussions

The designed TVLW comprises three domains A |B$_x$| C (see **Figures 1**a and 1b), where $x$ represents the number of layers in domain B. Domain B is a PhC with lattice constant $a$ = 12.12 mm, $w$ = 0.6 mm, $r$ = 7 mm, and $l_1$ = $l_2$ = 2.5 mm. Each unit cell has a metallic pattern with 35-μm thickness, deposited on a dielectric substrate (F4B with relative permittivity $\varepsilon$ = 2.20 and thickness $h$ = 2 mm). Due to the $C_6$ symmetry, the PhC in domain B features a pair of Dirac points at K and K' valleys in the band structure at 9.39 GHz (see blue lines in Figure 1c). Domain A and C are two valley PhCs with opposite valley-Chern numbers, which have similar unit cells to those in domain B, except that $l_1$ = 2.1 mm and $l_2$ = 2.9 mm in domain A and $l_1$ = 2.9 mm and $l_2$ = 2.1 mm in domain C. Because $l_1 \neq l_2$, the inversion symmetry of the PhCs in domain A and C is

broken. Consequently, the Dirac points are lifted, and a topological valley bandgap opens from 8.62 to 10.26 GHz (see red lines in Figure 1c).

Figure 1d shows the dispersion relations of the guided modes in the TVLW with layers $x = 8$, i.e., A |$B_8$| C. One can see a gapless dispersion marked as a red line within the bulk bandgap. The group velocity of the topological guided mode bounded to K and K' valleys is exactly opposite, exhibiting the valley-chirality locking property, which shares similarity to that of valley kink states in A | C domain wall.[5-15] From the simulated field ($H_z$) map of guided mode (see the right panel of Figure 1d), one can see the field concentrates in entire domain B and decays into domain A and C, in contrast to the field distributions of the valley kink states. In addition, there are two $0^{th}$ non-topological guided modes with gapped dispersions that lack momentum-valley locking property.[16] Between $0^{+th}$ and $0^{-th}$ modes, there is a topological frequency window highlighted in shaded grey, where only the topological guided mode exists. In Figure 1e, it is observed that the width of the topological frequency window decreases as $x$ in domain B increases. Meanwhile, higher-order non-topological guided modes appear in the bulk bandgap. To understand it, one can consider an extreme case that, when $x$ in domain B is large enough, the bulk-edge correspondence between domain A and C is so weak that the topological frequency window tends to close to finally reproduce the Dirac points of the PhC in domain B.

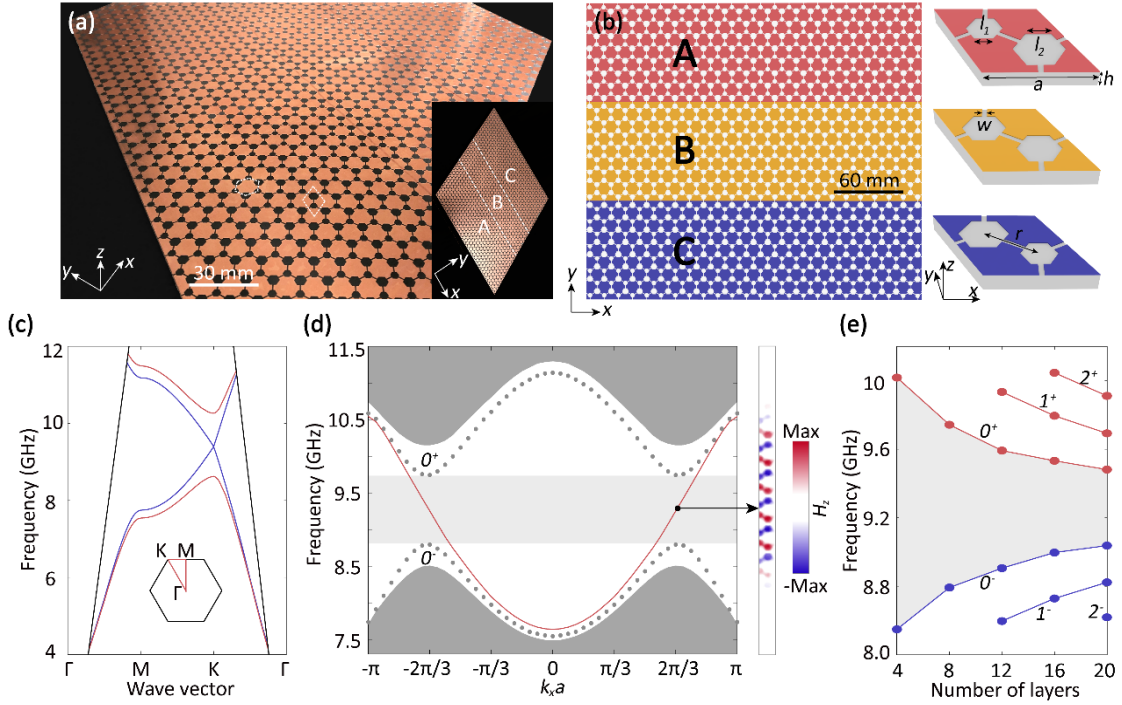

**Figure 1. A topological valley-locked waveguide (TVLW) consisting of three domains A |$B_x$| C. (a)** Photograph of an experimental sample. The white rhombus or the hexagon represents a unit cell. **(b)** Schematic view of the TVLW. Right panel: unit cells of PhCs in domain A and C with broken inversion symmetry, and that in domain B with inversion symmetry. The structural parameters are $l_1$ = 2.1 mm and $l_2$ = 2.9 mm for domain A ($l_1$ = $l_2$ = 2.5 mm for domain B, $l_1$ = 2.9 mm and $l_2$ = 2.1 mm for domain C), $h$ = 2 mm, $w$ = 0.6 mm, $r$ = 7 mm, and lattice constant $a$ = 12.12 mm. The background dielectric material has a relative permittivity $\varepsilon$ = 2.20. **(c)** Band diagrams of PhC in domain B (blue lines) with a Dirac point at 9.39 GHz, and band diagrams of PhC in domain A/C (red lines) with a bandgap from 8.62 to 10.26 GHz. Inset is the first Brillouin zone. **(d)** Band diagrams of A |$B_8$| C domain wall. Red line: a topological guided mode. The 0$^{th}$ bands are non-topological gapped guided modes. The grey regions represent the projection of the bulk bands. The shaded area depicts a topological

frequency window, where only the topological guided mode exists. Right panel: simulated field map of $H_z$ component showing the topological guided mode. **(e)** The width of the topological frequency window as a function of number of layers $x$ in domain B.

The topological guided mode in the TVLW can be explained from an effective Hamiltonian model intuitively. According to the $k \cdot p$ theory,[7,29] the effective Hamiltonian of the PhCs around the K valley is

$$\delta H_K(\delta k) = v_D \delta k_x \sigma_x + v_D \delta k_y \sigma_y + m v_D^2 \sigma_z$$

where $\delta k = k - k_K$ is the displacement of wave vector $k$ to K valley in momentum space, $v_D$ is the group velocity, and $\sigma_i = (i=x, y, z)$ are elements in the Pauli matrices. $m$ is the effective mass term with $m < 0$ for domain A, $m = 0$ for domain B, and $m > 0$ for domain C. We can also obtain the effective Hamiltonian around K' valley by applying the time-reversal operation. Solving the eigenvalue equation $\delta H_K = \delta w \phi$, the dispersion relation can be derived as $\delta^2 w = v_D^2 (\delta^2 k_x + \delta^2 k_y) + m^2 v_D^4$. Next, we consider the topological guided mode $\phi_{ABC}$ in the TVLW. According to the field map in Figure 1d that the $\phi_{ABC}$ mode exponentially attenuates along $+y$ direction in domain A and along $-y$ in domain C, we obtain a specific solution with $\delta w = v_D \delta k_x$ and $\phi_{ABC} = (1,1)^T$, whose slope $v_D$ is the same as that of bulk states in domain B ($\delta w = \pm v_D \delta k$). Thus, the topological guided mode could be regarded as a combination of the valley kink states in A | C domain wall and the bulk states in domain B.[16]

We then plot the simulated $H_z$ field maps of three TVLWs with $x = 2, 8, 16$ at 9.20

GHz in **Figure 2**a. As displayed in the field maps, the energy is concentrated in domain B for all three cases. Besides, by integrating $H_z$ field on the waveguide cross section, we obtain the energy capacity for the TVLWs with different $x$, which obviously reveals that a TVLW with larger $x$ in domain B has higher capacity for energy transport (i.e., 9.3 A•m for $x = 2$, 33.0 A•m for $x = 8$, and 61.4 A•m for $x = 16$). Note that we use a line of 2, 8, 16 identical point sources for TVLWs with $x = 2, 8, 16$, respectively, in order to guarantee the uniform energy distributions along the waveguides.

Next, we set an experiment to visualize the topological guided mode in a straight TVLW with $x = 8$. In the experiments, an electric-dipole-like antenna is fixed on the left side of the TVLW, to excite the topological guided mode, and another magnetic dipole antenna is placed above the sample to measure the field patterns or transmissions. Both antennas are connected to the vector network analyzer (VNA) to record the amplitude and phase of the measured field (see more details in Supporting Information).

The measured transmissions (see Figure 2b) show an obvious bandgap from 8.60 to 10.28 GHz. Moreover, we directly map out the topological guided mode by scanning the $H_z$ component on the $xy$ plane above the experimental sample. Figure 2c is the measured $|H_z|^2$ field distributions at 9.60 GHz, which clearly shows the energy is localized in domain B with suppressed inter-valley scattering. The spatial Fourier spectrum in the inset of Figure 2c (transformed from the interior region marked as a green rectangle) indicates that only the K valley is excited, hence the topological guided mode is K-valley locked. Besides, by applying the spatial Fourier transform along the green dashed line in Figure 2c, we obtain the experimental dispersion relations of the

topological guided mode (see Figure 2d), which have a good agreement with the simulated counterparts.[7,16]

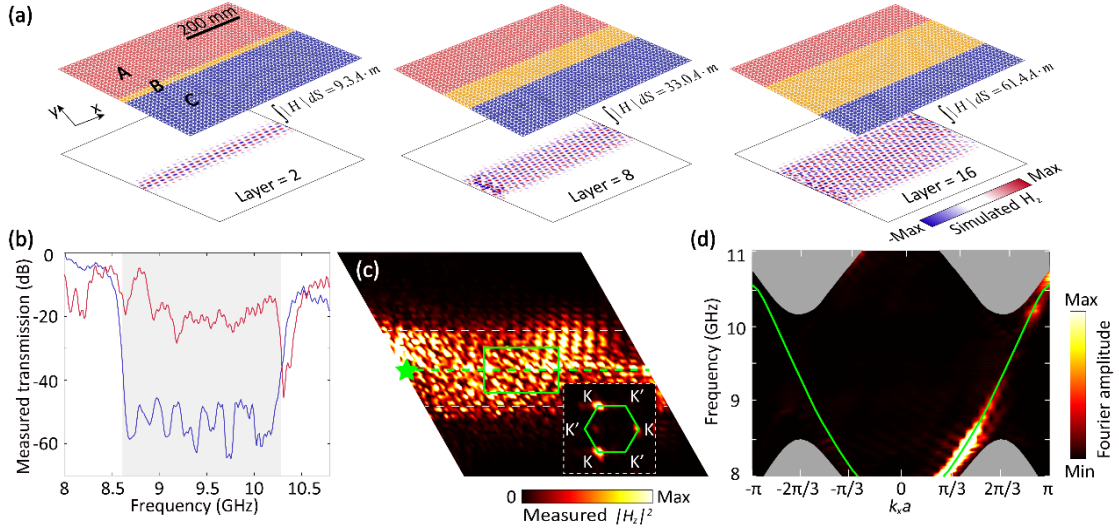

**Figure 2. Topological guided mode in a straight TVLW and its experimental characterization. (a)** Top panels: schematics of three straight TVLWs with $x$ = 2, 8, 16. Bottom panels: simulated $H_z$ field maps at 9.20 GHz in three straight TVLWs. **(b)** Measured transmissions in the straight TVLW with $x$ = 8. Blue line: bulk state. Red line: guided mode in the TVLW. Shaded area: bulk bandgap. **(c)** Measured $|H_z|^2$ field distributions at 9.60 GHz in the straight TVLW with $x$ = 8. Green star: point source. Inset: spatial Fourier spectrum transformed from the field pattern in the green rectangular area. Green hexagon: the first Brillouin zone. **(d)** Measured dispersions that are obtained by applying Fourier transform to the field pattern along a green dashed line in (c). Green solid lines are the simulated dispersions.

We further design a valley-locked topological channel intersection, based on the TVLW.[16] The configurations of the topological channel intersection are shown in the top panel of **Figure 3**a or 3b, which consists of five domains, and domain B has layers

$x = 8$. There are four ports labeled as 1-4 at the terminals of the outer domain walls. To experimentally show the device performance, we place the source at port 1 and port 4, respectively, and measure the field distributions at 9.60 GHz. The measured results are shown in the bottom panels of Figures 3a and 3b, where the green stars denote the sources. One can observe that the topological guided mode launched at port 1 (port 4) could propagate to the port 2 and port 4 (port 1 and port 3), and is suppressed when transporting to port 3 (port 2). This is because, when we excite the source at port 1 (port 4), the transport routes from port 1 to port 2 or port 4 (from port 4 to port 1 or port 3) have the same valley pseudo-spin, while the transport routes from port 1 to port 3 (from port 4 to port 2) have the opposite valley pseudo-spins.[7,14,16]

Besides, we measure the transmissions (i.e., $H_z$ amplitude ratios), which are labeled as $S_{ij}$ by the input port $j$ and output port $i$. From the transmissions, it is evident that the transmissions S21 and S41 are much larger than S31 within the photonic bandgap frequency window (see Figure 3c). Similar phenomenon could be found when launching the excitation at port 4 (see Figure 3d). These observations indicate that in the channels, the valley pseudo-spins are bounded to the propagations of the topological guided mode, confirming the momentum-valley locking in the TVLW.

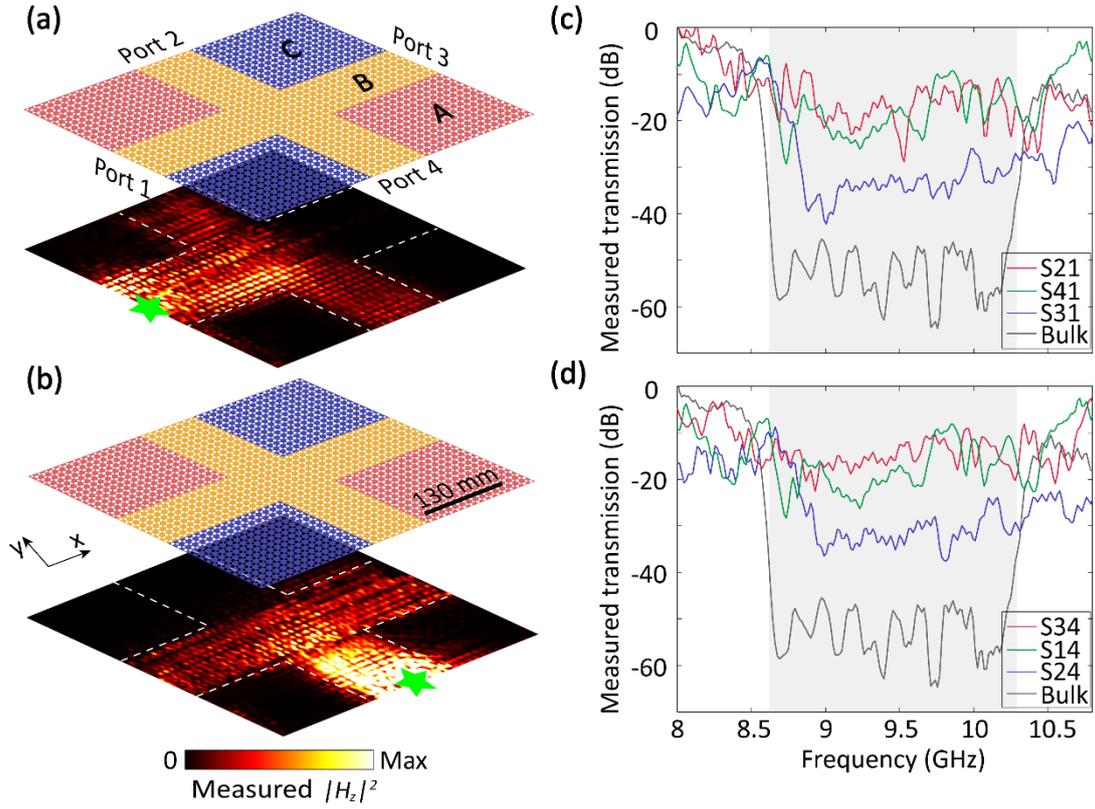

**Figure 3. A TVLW-based topological channel intersection. (a), (b)** Top panels: schematics of a topological channel intersection with layers $x = 8$. Bottom panels: measured $|H_z|^2$ field distributions in the topological channel intersection at 9.60 GHz when placing a point source at port 1 (a) and port 4 (b), respectively. Green star: point source. **(c), (d)** Measured transmissions when placing a point source at port 1 (c) and port 4 (d), respectively. The region highlighted in shaded gray indicates the bulk bandgap.

Since the existence of the topological guided mode is independent of the number of layers $x$ in domain B, we can take this advantage to realize valley-locked reflection-immune energy squeezing or concentrating.[16] **Figure 4**a shows the schematic of a topological concentrator composed of three domains. The width of domain B abruptly changes from $x = 16$ to $x = 1$ in the middle of the device. Figure 4b illustrates the experimental energy transport of the topological guided mode excited by a point source marked as a green star. It is evident that the energy firstly converges in the narrow part of domain B and then diverges in the broad part of domain B. To quantitatively analyze the energy intensity, the measured intensity profiles along three dashed green lines in Figure 4b are plotted in Figure 4c, where the energy intensity along line 2 is much larger than that along line 1 and line 3, indicating a strong field enhancement. Such a topological concentrator possibly finds applications in energy harvesting, field enhancement, and boosting nonlinear effect.[24-25]

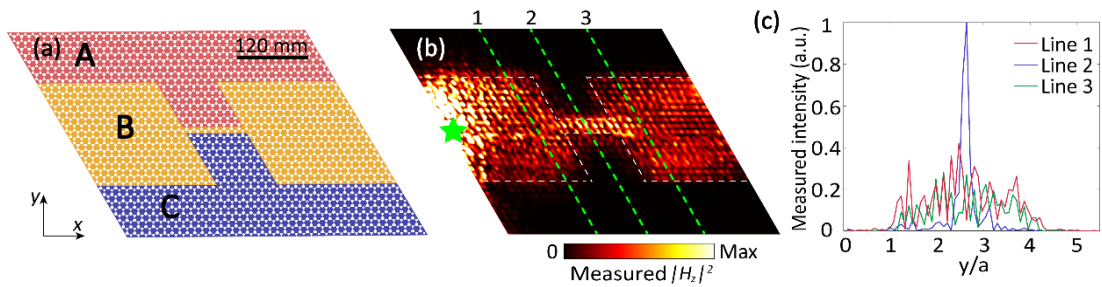

**Figure 4. A valley-locked topological energy concentrator.** **(a)** Schematic of a topological concentrator. The width of domain B abruptly changes from 16 layers to 1 layer in the middle of the device. **(b)** Measured $|H_z|^2$ field distributions in the topological concentrator at 9.55 GHz. Green star: point source. **(c)** Measured intensity profiles along three dashed green lines in (b).

Topological resonant cavities with tunable confinement and high energy capacity could also be devised based on the TVLW, which could be very useful for robust lasing.[17,26-28] Here, we design a topological cavity with triangular geometry that consists of A, C, B with layers $x = 4$, as depicted in the top panel of **Figure 5**a. The bottom panel of Figure 5a is the simulated $H_z$ field distributions in the cavity A |B$_4$| C at 8.9 GHz, showing the energy is uniformly distributed in the whole cavity. For reference, we design a topological cavity A | C with only A and C domains, and the corresponding simulated $H_z$ field distributions at 8.8 GHz are displayed in Figure 5b. Compared to the conventional topological cavity A | C,[17] the cavity A |B$_4$| C has higher energy capacity, which is evidenced by the simulated energy intensity along the dashed blue lines in Figures 5a and 5b, as shown in Figures 5c and 5d. Besides, the energy capacity is tailorable, by changing the number of layers $x$ in domain B.

To further confirm the topological cavity modes, we perform the experiments to measure the $|H_z|^2$ field distributions at 9.55 GHz (see Figure 5e). Additionally, the spectrum intensity of the topological cavity A |B$_4$| C illustrates four regularly spaced peaks of topological cavity modes, locating at 9.22, 9.55, 9.91, and 10.31 GHz (see Figure 5f). These regularly spaced peaks are the signature of running modes circulating around the triangular cavity without experiencing localization, owing to the existence of topological protection of the cavity modes. Such a topologically-protected cavity with tunable confinement and high energy capacity could be very promising for novel topological lasing.[17,26-28]

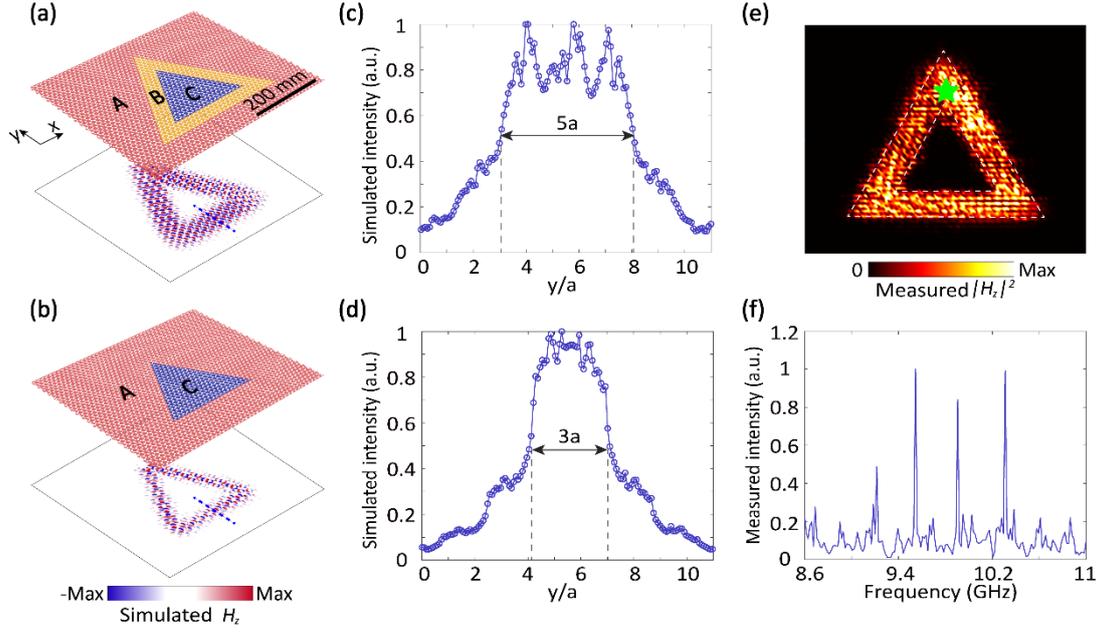

**Figure 5. A topological cavity based on the TVLW. (a), (b)** Top panels: schematics of topological cavities composed of A |B$_4$| C (a) and A | C (b), respectively. Bottom panels: simulated $H_z$ field distributions in the cavity A |B$_4$| C at 8.9 GHz (a) and in the cavity A | C at 8.8 GHz (b), respectively. **(c), (d)** Simulated energy intensity along the dashed blue lines in (a) and (b), respectively. **(e)** Measured $|H_z|^2$ field distributions in the cavity A |B$_4$| C at 9.55 GHz. Green star: point source. **(f)** Spectrum intensity of the topological cavity A |B$_4$| C with four regularly spaced peaks at 9.22, 9.55, 9.91, and 10.31 GHz.

## 3. Conclusions:

To conclude, we experimentally demonstrate a photonic TVLW where a Dirac PhC is sandwiched by two valley PhCs with opposite valley-Chern numbers. The topological guided mode in the TVLW is characterized with gapless dispersions, momentum-valley locking, robustness against defects, and tunable mode width. The present TVLW has found some unique applications, such as high-energy-capacity topological channel intersections, valley-locked energy concentrators, and topological cavities with tailorable mode confinement, which are enabled by the mode width DOF. Besides, in comparison to the previous domain wall structures hosting the valley kink states, the TVLW is more flexible to interface with the existing photonic waveguides and devices. Looking ahead, there are further potentials to incorporate the TVLW with mode width DOF into other photonic active device applications with high energy capacities.

**Supporting Information**

Supporting Information is available from the Wiley Online Library or from the author.

**Acknowledgements**

The work at Zhejiang University was sponsored by the National Natural Science Foundation of China (NNSFC) under Grants No. 61625502, No.11961141010, and No. 61975176, the Top-Notch Young Talents Program of China, and the Fundamental Research Funds for the Central Universities.

**Conflict of Interest**

The authors declare no conflict of interest.

**References:**
[1] Y. P. Shkolnikov, E. P. De Poortere, E. Tutuc, M. Shayegan, *Phys. Rev. Lett.* **2002**, 89.
[2] O. Gunawan, Y. P. Shkolnikov, K. Vakili, T. Gokmen, E. P. De Poortere, M. Shayegan, *Phys. Rev.*

*Lett.* **2006**, 97.

[3] A. Rycerz, J. Tworzydlo, C. W. J. Beenakker, *Nat. Phys.* **2007**, 172.

[4] J. R. Schaibley, H. Yu, G. Clark, P. Rivera, J. S. Ross, K. L. Seyler, W. Yao, X. Xu, *Nat. Rev. Mater.* **2016**, 1.

[5] T. Ma, G. Shvets, *New J. Phys.* **2016**, 25012.

[6] Z. Gao, Z. Yang, F. Gao, H. Xue, Y. Yang, J. Dong, B. Zhang, *Phys. Rev. B* **2017**, 96.

[7] X. Wu, Y. Meng, J. Tian, Y. Huang, H. Xiang, D. Han, W. Wen, *Nat. Commun.* **2017**, 8.

[8] F. Gao, H. Xue, Z. Yang, K. Lai, Y. Yu, X. Lin, Y. Chong, G. Shvets, B. Zhang, *Nat. Phys.* **2018**, 14.

[9] J. Noh, S. Huang, K. P. Chen, M. C. Rechtsman, *Phys. Rev. Lett.* **2018**, 120.

[10] X. Chen, W. Deng, J. Lu, J. Dong, *Phys. Rev. B* **2018**, 97.

[11] M. I. Shalaev, W. Walasik, A. Tsukernik, Y. Xu, N. M. Litchinitser, *Nat. Nanotechnol.* **2019**, 14.

[12] X. He, E. Liang, J. Yuan, H. Qiu, X. Chen, F. Zhao, J. Dong, *Nat. Commun.* **2019**, 10.

[13] Q. Chen, L. Zhang, M. He, Z. Wang, X. Lin, F. Gao, Y. Yang, B. Zhang, H. Chen, *Adv. Opt. Mater.* **2019**, 7.

[14] L. Zhang, Y. Yang, M. He, H. X. Wang, Z. Yang, E. Li, F. Gao, B. Zhang, R. Singh, J. H. Jiang, H. Chen, *Laser Photonics Rev.* **2019**, 13.

[15] Y. Yang, Y. Yamagami, X. Yu, P. Pitchappa, J. Webber, B. Zhang, M. Fujita, T. Nagatsuma, R. Singh, *Nat. Photon.* **2020**, 14.

[16] M. Wang, W. Zhou, L. Bi, C. Qiu, M. Ke, Z. Liu, *Nat. Commun.* **2020**, 11.

[17] Y. Zeng, U. Chattopadhyay, B. Zhu, B. Qiang, J. Li, Y. Jin, L. Li, A. G. Davies, E. H. Linfield, B. Zhang, Y. Chong, Q. J. Wang, *Nature* **2020**, 578.

[18] A. G. Rickrnan, G. T. Reed, *IEE Proc.-Optoelectron.* **1994**, 6.

[19] Y. Rong, K. A. Zaki, *IEEE T. Microw. Theory* **2000**, 48.

[20] O. Powell, *J. Lightwave Technol.* **2002**, 20.

[21] Y. Cassivi, L. Perregrini, P. Arcioni, M. Bressan, K. Wu, G. Conciauro, *IEEE Microw. Wirel. Co.* **2002**, 12.

[22] M. Silveirinha, N. Engheta, *Phys. Rev. Lett.* **2006**, 97.

[23] R. Liu, Q. Cheng, T. Hand, J. J. Mock, T. J. Cui, S. A. Cummer, D. R. Smith, *Phys. Rev. Lett.* **2008**, 100.

[24] Q. Liang, T. Wang, Z. Lu, Q. Sun, Y. Fu, W. Yu, *Adv. Opt. Mater.* **2013**, 1.

[25] M. Soljacic, J. D. Joannopoulos, *Nat. Mater.* **2004**, 3.

[26] B. Bahari, A. Ndao, F. Vallini, A. A. El, Y. Fainman, B. Kante, *Science* **2017**, 358.

[27] G. Harari, M. A. Bandres, Y. Lumer, M. C. Rechtsman, Y. D. Chong, M. Khajavikhan, D. N. Christodoulides, M. Segev, *Science* **2018**, 359.

[28] M. A. Bandres, S. Wittek, G. Harari, M. Parto, J. Ren, M. Segev, D. N. Christodoulides, M. Khajavikhan, *Science* **2018**, 1231.

[29] J. Mei, Y. Wu, C. T. Chan, Z. Zhang, *Phys. Rev. B*, **2012**, 86.